\documentclass[prd,amsmath,amsfonts,preprint,11pt,superscriptaddress,onecolumn]{revtex4}
\usepackage[dvips]{color}
\usepackage{amsmath}
\usepackage{amsfonts}
\usepackage{amssymb}

\usepackage{graphics}
\usepackage{amsmath}
\usepackage{amsfonts}
\usepackage{amssymb}
\usepackage{graphicx}
\usepackage{epsfig}

\def\ba{\begin{eqnarray}}
\def\ea{\end{eqnarray}}
\def\be{\begin{equation}}
\def\ee{\end{equation}}

\def\nn{\nonumber}

\def\({\left(}
\def\){\right)}

\def\be{\begin{equation}}
\def\ee{\end{equation}}
\def\bea{\begin{eqnarray}}
\def\eea{\end{eqnarray}}
\def\nn{\nonumber}
\def\ansatz{{\it ansatz}}
\def\ansatze{{\it ans\"atze}}
\def\exd{{\rm d}}
\def\pref#1{(\ref{#1})}
\def\cA{{\cal A}}

\begin{document}

%%CB: Groping around here for a good title....

%\title{Notes on dS analogues of GGP solutions}
\title{Bulk Singularities and the Effective Cosmological Constant\\
for Higher Co-dimension Branes}

\author{Andrew J. Tolley}
\email{atolley@princeton.edu} \affiliation{Joseph Henry
Laboratories, Princeton University, Princeton NJ, 08544, USA.}

\author{C.P. Burgess}\email{cburgess@perimeterinstitute.ca}
\affiliation{Department of Physics and Astronomy, McMaster
University, Hamilton, ON, L8S 4M1, Canada,\\
and Perimeter Institute, Waterloo, ON, N2L 2Y5, Canada.}

\author{D. Hoover}\email{doug.hoover@mail.mcgill.ca}
\affiliation{Physics Department, McGill University, Montr\'eal,
Qu\'ebec, H3A 2T8, Canada.}

\author{Y. Aghababaie}

\date{\today}

\begin{abstract}
We study a general configuration of parallel branes having
co-dimension $\ge 2$ situated inside a compact $d$-dimensional
bulk space within the framework of a scalar and flux field coupled
to gravity in $D$ dimensions, such as arises in the bosonic part
of some $D$-dimensional supergravities.
A general relation is derived which relates the induced curvature
of the observable noncompact $n$ dimensions to the asymptotic
behaviour of the bulk fields near the brane positions. For
compactifications down to $n = D-d$ dimensions we explicitly solve
the bulk field equations to obtain the near-brane asymptotics, and
by so doing relate the $n$-dimensional induced curvature to
physical near-brane properties. In the special case where the bulk
geometry remains nonsingular (or only conically singular) at the
brane positions our analysis shows that the resulting $n$
dimensions must be flat. As an application of these results we
specialize to $n=4$ and $D=6$ and derive a new class of solutions
to chiral 6D supergravity for which the noncompact 4 dimensions
have de Sitter or anti-de Sitter geometry.
\end{abstract}
\maketitle
\tableofcontents
%\newpage

\section{Introduction}

In four dimensions the twin requirements of general covariance and
the Lorentz-invariance of the vacuum imply that the vacuum energy
inevitably appears to gravity like a 4D cosmological constant,
with a vacuum energy, $\rho$, corresponding to a cosmological
constant of order $\Lambda = 8 \pi G \rho$ (where $G$ here denotes
Newton's constant). The cosmological constant problem
\cite{ccreview} refers to the huge mismatch between the large
vacuum energy expected from the known quantum zero-point
fluctuations and the very small upper limit on (or the observed
value for) the cosmological constant coming from cosmology.

Higher dimensional theories are of interest for the cosmological
constant problem because they offer the possibility that the
gravitational influence of a 4D vacuum energy need not be a 4D
cosmological constant. In particular, within a higher-dimensional
context the possibility exists that the gravitational response of
a large 4D vacuum energy might be to curve the extra dimensions
rather than the observable four, raising the hope that a large
vacuum energy need not lead to a large 4D cosmological constant.
The introduction of branes into the picture considerably sharpens
this hope, since solutions exist to the higher-dimensional field
equations for which the effective 4D cosmological constant
vanishes even though they are sourced by large 4D energy
configurations (typically large brane tensions).

In recent years these observations have stimulated several
proposals to realize this possibility in a concrete way
\cite{precursors,codim1,carroll,sled}, all with the theme that
large 4D energy densities need not imply a strongly-curved 4D
geometry within a brane-world picture. Although this is arguably a
step forward, it is not the end of the story since the mere
existence of such solutions does not directly address the issues
of fine-tuning which underly the cosmological constant problem.
These issues come in several forms, either to do with the
stability of the solution under the quantum renormalization of the
underlying parameters, or to do with stability of the time
evolution of the solutions against perturbations in the initial
conditions (for a review of some of these concerns see
\cite{TAMU}).

One of the key questions underlying these naturalness issues asks:
What conditions must be required of the various source brane
configurations in order to make the observed 4 dimensions flat?
This question is crucial for addressing the fine-tuning issue
because one must always be on guard against hidden fine tunings.
In particular, if the properties of various branes must be
carefully adjusted (or adjusted relative to one another) then it
is the stability of {\it this} particular adjustment (against
renormalization, say) which must be established in order to solve
the cosmological constant problem. Indeed the main criticisms to
proposals \cite{precursors,codim1,carroll,sled} fall into this
category \cite{codim1x,carrollx,sledx} (see also
\cite{TAMU,update}).

It is our purpose in this paper to provide a general answer to
this question for a scalar-tensor-flux field equations arising
in $D$-dimensional supergravity theories, for solutions having $n$
maximally-symmetric dimensions that are sourced by branes having
co-dimension $\ge 2$.
(The case of co-dimension 1 -- as appropriate for Randall-Sundrum
models \cite{RS}, for instance -- differs from other co-dimensions
and is presently better understood \cite{codim1x}.) We defer to a
later paper the discussion of the naturalness issues associated
with the quantum corrections to, and the stability of, the
solutions presented here.

For these systems we obtain the following results:
\begin{itemize}
\item We derive a general expression,
eq.~\pref{Curvature-BraneAsympt}, which relates (a particular
average over the extra dimensions of) the curvature of the
maximally-symmetric $n$ dimensions to the asymptotic form taken by
the bulk metric very close to the source branes. Our expression
generalizes similar expressions which have been derived, either
for 6D supergravities in the co-dimension 2 case \cite{GGP} or for
higher-dimensional non-supersymmetric gravity \cite{ML}. Our
result also applies to FRW-like time-dependent geometries for
which the $n$ maximally-symmetric dimensions are spatial, in which
case the spatial curvature is related to both the near-brane
asymptotic forms and to contributions from spatial slices in the
remote past and future.
\item We provide a very general classification of the near-brane
form taken by the bulk fields near their sources. Using arguments
in the spirit of the BKL analysis of time-dependence near
singularities \cite{BKL} we show that in the near-brane limit the
higher-dimensional supergravity fields have a power-law dependence
on the proper distance, $r$, from the branes. We show that the
bulk fields are very generically singular near the branes, and
that the bulk field equations impose Kasner-like relations,
eqs.~\pref{constraint1} and \pref{constraint2}, amongst these
powers, which strongly restrict the kinds of powers (and so also
the singularities) which arise.
\item Combining the above two points allows an identification of
how the curvature of the large $n$ dimensions depends on the
asymptotic powers which govern the asymptotic near-brane behaviour
of the bulk fields. This relation shows that the large $n$
dimensions must be flat in the absence of singularities within the
extra dimensions (or if these singularities are only conical). In
the more generic case of singular configurations we find that a
flat $n$ dimensions requires either the extra-dimensional warp
factor, $W$, or the dilaton, $e^\varphi$, must grow like an
inverse power of $r$ as the brane is approached ({\it i.e.} as $r
\to 0$). (In our conventions $e^\varphi \ll 1$ corresponds to weak
coupling in string theory.)
\item As an application we specialize the above results to the
case of 6D supergravity compactified to 4 dimensions, and use them
to show the existence of a new class of solutions for which the
maximally-symmetric 4 dimensions are de Sitter-like (or anti-de
Sitter-like), unlike all of those which are presently known.
\end{itemize}

Our presentation is organized in the following way. The next
section sets up the supergravity equations of interest and their
compactification to $n$ maximally-symmetric dimensions. It is here
that we derive the key relationship,
eq.~\pref{Curvature-BraneAsympt}, relating the $n$-dimensional
curvature to the asymptotics of bulk fields near the source-brane
singularities. Section III then examines the relevant near-brane
asymptotic forms for the bulk fields, and derives the power-law
behaviour which the bulk equations dictate. These are then used in
the results of Section II to more directly relate the
$n$-dimensional curvature to the power-law dependence of the bulk
fields in the near-brane limit. Finally, Section IV specializes to
6D supergravity compactified to 4 maximally-symmetric dimensions,
and shows how to use the previous two sections to generalize the
class of 6D solutions to include those having de Sitter-like and
anti-de Sitter-like 4-dimensional slices.

\section{The Curvature-Asymptotics Connection}

In this section we summarize the field equations of interest,
which are the bosonic parts of the equations of motion for many
higher-dimensional supergravities. We also here specialize the
fields appearing in these equations to the most general
configurations which are maximally symmetric in (3+1) non-compact
dimensions, as is appropriate for describing the warped
compactifications of interest. We allow these solutions to have
singularities (more about which below) at various points within
the extra dimensions corresponding to the positions of various
branes having co-dimension $\ge 2$. Our goal in so doing is to
establish a general connection, eq.~\pref{Curvature-BraneAsympt},
between the curvature of the noncompact 4D geometry and the
asymptotic behaviour of the bulk fields in the vicinity of the
various branes.

\subsection{The Field Equations}

Our starting point is the following action in $D$ spacetime
dimensions
\be
    \label{DDAction}
    S = -\int \exd^Dx \sqrt{-g} \left[ \frac{1}{2 \kappa^2} \,
    g^{MN} \Bigl( R_{MN} + \partial_M \varphi \,
    \partial_N \varphi \Bigr) + \frac{1}{2} \sum_r
    \frac{1}{(p_r+1)!} e^{-p_r \varphi} F_{r}^2
    + \cA \, e^{\varphi} \right] \,,
\ee
where $\kappa^2 = 8 \pi G$ denotes the higher-dimensional Newton
constant and $\cA$ is a dimensional constant. The fields $F_{r}$
are the $(p_r+1)$-form field strengths for a collection of
$p_r$-form gauge potentials, $A_{r}$, and $F^2 =
F_{M_1..M_{p_r+1}} F^{M_1..M_{p_r+1}}$. When $\cA = 0$ this is
sufficiently general to encompass the bosonic parts of a variety
of higher-dimensional, ungauged supergravity lagrangian densities
\cite{HiDSugra}. When $\cA \ne 0$ the dilaton potential has the
form found in chiral 6D supergravity \cite{NS}.

The field equations obtained from this action are:
\bea \label{einstein}
    &&\Box \, \varphi  - \kappa^2 \cA \, e^{\varphi} + \kappa^2\sum_r
    \frac{p_r}{2(p_r+1)!} e^{-p_r \varphi} F_{r}^2 = 0
     \qquad \hbox{(dilaton)}\nn \\
    &&\nabla_M \Bigl(e^{ -p_r \varphi} \, F_{r}^{MN \ldots Q} \Bigr)
    + (\hbox{CS terms})= 0 \qquad
 \hbox{($p_r$-form)}\\
    &&R_{MN} + \partial_M\varphi \, \partial_N\varphi
    + \kappa^2\sum_r \frac{1}{p_r!} e^{-p_r \varphi} \left[ F_{r}^2
    \right]_{MN} + \frac{2}{D-2} \, (\Box\,
    \varphi)\, g_{MN} =0 \qquad \hbox{(Einstein)}
    \,,\nn
\eea
where `(CS terms)' denotes terms arising from any Chern-Simons
terms within the definition of $F_{(r)}$, and we define
\be
    \left[ F^2 \right]_{MN} = F^{\hspace{7pt} P \ldots R}_M
    F_{NP \ldots R} \,.
\ee
The ability to write the term proportional to $g_{MN}$ in the
Einstein equation in terms of $\Box \varphi$ is a consequence of
the particular powers of $e^\varphi$ which pre-multiply each of
the terms in the action, \pref{DDAction}. This choice corresponds
to the existence of a scaling symmetry of the classical field
equations, according to which
\be
    g_{MN} \to \omega \, g_{MN} \quad \hbox{and} \quad
    e^\varphi \to \omega^{-1} \, e^\varphi \,,
\ee
with constant $\omega$ and the field strengths, $F_r$, not
transforming. Although this is not a symmetry of the action, which
transforms as $S \to \omega^{(D-2)/2} S$, it does take solutions
of the classical equations into one another.

\subsection{Maximally-Symmetric Compactifications}

We seek solutions to these equations for which $n$ dimensions are
maximally symmetric and $d=D-n$ are not. In most applications we
have in mind $n=4$, corresponding to having 3+1
maximally-symmetric directions and $d=D-4$ static, compact
euclidean dimensions. But our analysis is general enough also to
include (with minor modifications) situations of interest to
cosmology for which there are $n=3$ maximally-symmetric spatial
dimensions and $d=D-4$ time-dependent, compact dimensions.

To this end divide the $D$ coordinates $x^M$, $M = 1...D$, into
$n$ maximally-symmetric coordinates, $x^\mu$, $\mu = 1...n$, and
the remaining $d=D-n$ coordinates, $y^i$, $i = n+1...d$. We use
the metric \ansatz\ which follows from maximal symmetry:
\be
    \label{metricansatz}
    \exd s^2 = \hat{g}_{MN} \, \exd x^M \, \exd x^N
    = W^2(y) \, g_{\mu\nu}(x) \, \exd x^\mu \, \exd x^\nu +
    \tilde{g}_{ij}(y) \, \exd y^i \exd y^j \,,
\ee
where $g_{\mu\nu}$ is an $n$-dimensional maximally symmetric
metric and $\tilde{g}_{ij}$ a generic $d$-dimensional metric.
Throughout this section, we use the convention that hats denote
objects constructed from the full $D$-dimenional metric
$\hat{g}_{MN}$, while tildes denote objects constructed from the
metric $\tilde{g}_{ij}$. Tensors without hats or tildes are
constructed from the metric $g_{\mu\nu}$.

With these conventions the Einstein equation, eq.~\pref{einstein},
specialized to the maximally-symmetric directions reads
\be
  \label{einstein2}
  \hat{R}_{\mu\nu} + \frac{2}{D-2} (\hat{\Box} \varphi) \hat{g}_{\mu\nu}
  =0\,,
\ee
where we use that maximal symmetry implies $\partial_\mu \varphi =
0$ and $F_r^{\mu N..P} = 0$ (and so $\left[ F^2_r \right]_{\mu\nu}
= 0$).

\subsection{Relating Curvature to Bulk Asymptotics}

Using the metric \ansatz, \pref{metricansatz}, we may write
\be
    \hat{g}_{\mu\nu} = W^2 g_{\mu\nu} \,, \qquad
    \hat{R}_{\mu\nu} = R_{\mu\nu} + \frac{1}{n} (W^{2-n}
    \tilde{\nabla}^2 W^n) \, g_{\mu\nu} \quad \hbox{and} \quad
    \hat{\Box} \varphi = W^{-n} \tilde{\nabla}_i
    ( W^n \tilde{g}^{ij} \partial_j \varphi) \,,
\ee
where $\tilde{\nabla}^2 = \tilde{g}^{ij} \tilde\nabla_i
\tilde\nabla_j$. Since maximal symmetry implies $R_{\mu\nu} =
(R/n) \, g_{\mu\nu}$, these equations allow eq.~\pref{einstein2}
to be simplified to
\be \label{Curvature-BraneAsympt}
    \frac{1}{n} \, W^{n-2} \, R = -\tilde{\nabla}_i
    \left[ W^n \tilde{g}^{ij} \partial_j \left(\,
    \ln W + \frac{2\, \varphi}{D-2} \right) \right] \,.
\ee
This last equation represents the main result of this section, and
is a generalization to arbitrary dimensions of a similar result in
6 dimensions derived in ref.~\cite{GGP}.

The significance of eq.~\pref{Curvature-BraneAsympt} is most
easily seen once it is integrated over the compact $d$ dimensions
and Gauss' Law is used to rewrite the right-hand side as a surface
term:
\be \label{IntegratedCBA}
    \frac{1}{n} \int_M \exd^{d}y \; \sqrt{\tilde{g}}
    \, W^{n-2} \, R = -\sum_\alpha \int_{\Sigma_\alpha} \exd^{d-1}y
    \; \sqrt{\tilde{g}} \,
    {N}_i \left[ W^n \tilde{g}^{ij} \partial_j \left(\,
    \ln W + \frac{2\, \varphi}{D-2} \right) \right] \,,
\ee
where ${N}_i$ is an outward-pointing normal, with $\tilde{g}^{ij}
{N}_i {N}_j = 1$. (If time is one of the $d$ dimensions then the
surface terms must include spacelike surfaces in the remote future
and past, for which $\tilde{g}^{ij} N_i N_j = -1$.) If there are
no singularities or boundaries in the dimensions being integrated
then the right-hand side vanishes, leading to the conclusion that
the product $W^{n-2} \, R$ integrates to zero. Since $R$ is
constant and $W^{n-2}$ is strictly positive, this immediately
implies $R=0$, as concluded for 6D in ref.~\cite{GGP}.

Our interest in what follows is the case where the right-hand side
of eq.~\pref{Curvature-BraneAsympt} {\it does} have singularities
corresponding to the presence of various source branes situated
throughout the extra dimensions. In this case
eq.~\pref{Curvature-BraneAsympt} still carries content provided we
excise a small volume about the positions of each singularity,
thereby leaving a co-dimension-1 boundary, $\Sigma_\alpha$, which
surrounds each of the various brane positions. In this case
eq.~\pref{IntegratedCBA} directly relates the curvature of the
maximally-symmetric $d$ dimensions to the sum over the
contributions to the right-hand side of the boundary contributions
from each surface $\Sigma_\alpha$. Since these surfaces are chosen
to be close to the source branes, these surface contributions can
be simplified using the asymptotic forms taken by the bulk fields
in the immediate vicinity of these sources. After a brief
digression concerning the possible existence of horizons in these
geometries, we return in the next section to identify what these
asymptotic forms must be.

\subsubsection*{Horizon formation}

It is possible that for certain choices of brane sources horizons
form at some finite proper distance from the branes
\cite{GerochTraschen}. We investigate here the situations for when
this can occur, since such horizons could have implications for
the crucial sum rule, eq.~(\ref{IntegratedCBA}). We consider three
possible cases:

\begin{enumerate}
\item If $n > 1$ and one of the $x$ coordinates is time, then the
assumption of maximal symmetry implies the metric, $g_{\mu\nu}$,
has either an $ISO(n-1,1)$, $SO(n,1)$ or $SO(n-1,2)$ isometry
group. Such a symmetry group precludes the formation of horizons.
\item If $n > 1$ and one of the $y$ coordinates is time, then a
horizon could be present in the bulk, but this does not in itself
interfere with the validity of the above formula
(\ref{IntegratedCBA}). Rather, it might instead imply that both
spacelike and timelike boundaries will contribute on its RHS.
\item In the special case $n=1$ we have $R=0$ trivially, leading
to the stronger statement that the divergence on the RHS of
\pref{IntegratedCBA} vanishes. In this case the sum rule
(\ref{IntegratedCBA}) breaks down in the presence of a horizon
because the signature of the $\tilde{g}$ metric changes at the
horizon and so Gauss' Law is no longer well defined.
\end{enumerate}

\section{Near-Brane Solutions}

In this section we identify the general asymptotic form taken by
the bulk fields in the immediate vicinity of any source branes,
with an eye to its use in eq.~\pref{IntegratedCBA} of the previous
section. We are able to keep our analysis quite general by arguing
that these asymptotic forms are given by powers of the distance
from the source for co-dimension $> 2$ (or possibly logs for
co-dimension 2)
with the powers determined by explicitly solving the bulk
equations. Assuming these equations are dominated near the branes
by the contributions of the kinetic terms they may be integrated
quite generally, leading to solutions corresponding to Kasner-like
\cite{Kasner} near-brane geometries. Given these solutions the
validity of the assumption that kinetic terms dominate can be
checked {\it a posteori}. Our arguments closely resemble similar
arguments used long ago \cite{BKL} to identify the time-dependence
of spacetimes in the vicinity of space-like singularities.

\subsection{Asymptotic Near-Brane Geometries}

To this end we assume that the dilaton field, $\varphi$, and the
metric near the brane have the form
\be \label{dilatonandmetricform}
     \varphi \approx q \ln r   \quad \hbox{and} \quad
     \exd s^2 \approx r^{2w} \, g_{\mu\nu}(x) \, \exd x^\mu \, \exd x^\nu +
    \exd r^2 + r^{2\alpha} f_{ab}(z) \, \exd z^a \exd z^b\,,
\ee
where $w$, $\alpha$ and $q$ are constants. With respect to our initial
metric \ansatz, eq.~\pref{metricansatz}, we see that this
corresponds to the choices
\be
    \label{Wform}
    W(y) = r^{w} \qquad \hbox{and} \qquad
    \tilde{g}_{ij} \exd y^i \exd y^j = \exd r^2 + r^{2\alpha}
    f_{ab} \exd z^a \exd z^b,
\ee
where $\{y^i\} = \{r,z^a\}$. If the supergravity of interest is
regarded as describing the low-energy limit of a perturbative
string theory then our conventions are such that $e^\varphi \to 0$
represents the limit of weak string coupling. We see that if $q
<0$ then the region of small $r$ lies beyond the domain of the
weak-coupling approximation.

We imagine the brane location to be given by $r =0$ and the
coordinate $r$ is then seen to represent the proper distance away
from the brane. With this choice a surface having proper radius
$r$ has an area which varies with $r$ like $r^{\alpha (d-1)}$, and so
this area only {\it grows} with increasing $r$ if $\alpha > 0$. The
geometry in general has a curvature singularity at $r = 0$, except
for the special case $\alpha = 1$ for which the singularity can be
smooth (or purely conical).

Finally, we specialize for simplicity to the case where there is
only one non-vanishing gauge flux which we take to be for a
$p$-form potential whose field strength is $F$. With a
Freund-Rubin \ansatz\ \cite{FreundRubin} in mind we also
specialize to $p = d-1$ and take $F$ proportional to the volume
form of the $d$-dimensional metric $\tilde{g}_{ij}$. Near $r=0$,
we assume
\be \label{Fform}
     F^{r a_1 \dots a_p} \sim r^{\gamma}.
\ee

With these assumptions, we now determine the powers $\alpha$, $w$, $q$
and $\gamma$ by solving the field equations in the region $r
\approx 0$. We do so by neglecting the contributions of fluxes or
the dilaton potential in the dilaton and Einstein equations, and
by neglecting any Chern-Simons contributions to the equations for
the background $p$-form gauge potential. Once we find the
solutions we return to verify that the neglect of these terms is
indeed justified.

The $p$-form equation gives the condition
\be
     0 = \partial_r \Bigl( \sqrt{\hat{g}} e^{-p \varphi}
           F^{r z_1 \ldots z_p} \Bigr)
       \sim \partial_r \Bigl[ r^{wn+\alpha(d-1)-pq+\gamma} \Bigr]
\ee
which leads (when $p=d-1$) to the condition $\gamma =
(q-\alpha)(d-1)-wn$, and so
\be
     F^2 \sim r^{2q(d-1)-2wn} \,.
\ee

Consider next the dilaton equation. We first note that
\be
     \label{boxphi}
     \hat{\Box} \varphi = \frac{1}{\sqrt{\hat{g}}} \partial_M \left(
     \sqrt{\hat{g}} \, \hat{g}^{MN} \, \partial_N \varphi \right)
     \sim q [nw+\alpha(d-1)-1] \, r^{-2}.
\ee
For comparison, the other terms in the dilaton equation of motion
depend on $r$ as follows:
\be
     e^{-p \varphi} F^2 \sim r^{q(d-1)-2wn} \qquad \hbox{and}
     \qquad
     e^{\varphi} \sim r^q.
\ee
Thus, provided $q>-2$ and $q(d-1)-2wn>-2$ (whose domains of
validity we explore below) all of the terms in the dilaton
equation are subdominant to $\hat{\Box}\varphi$, and so may be
neglected. The dilaton therefore effectively satisfies $\hat\Box
\varphi = 0$ near $r=0$, and so from eq.~\pref{boxphi} we see that
this requires
\be
     \label{constraint1}
     nw+\alpha (d-1)=1.
\ee

Next consider the $rr$-component of the Einstein equation. Given
the assumed asymptotic form for the metric, we calculate
\bea
    \hat{R}_{rr} &=& [-wn + nw^2+ (\alpha^2-\alpha)(d-1) ] \,r^{-2} \nonumber \\
                 &=& [ n w^2+ \alpha^2(d-1) -1] \,r^{-2}.
\eea
As before, we find that the $F^2$ term is subdominant if
$q(d-1)-2wn > -2$. The
$rr$-Einstein equation therefore gives the additional constraint
\be
    \label{constraint2}
    n w^2+\alpha^2(d-1)+q^2=1.
\ee
Notice that this equation restricts the ranges of $w$, $\alpha$ and $q$
to be
\be \label{paramranges}
    -\frac{1}{\sqrt{n}} \le w \le \frac{1}{\sqrt{n}} \,,
    \quad
    - \frac{1}{\sqrt{d-1}} \le \alpha \le \frac{1}{\sqrt{d-1}} \,
    \quad \hbox{and} \quad
    -1 \le q \le 1 \,.
\ee
In particular it allows a regular solution (or one having a
conical singularity) -- {\it i.e.} one having $\alpha=1$ -- only if
$d=2$ and $q=w=0$.

The Einstein equations in the maximally symmetric dimensions can
be similarly evaluated using the assumed asymptotic form for the
metric. The contribution of the induced $n$-dimensional curvature
tensor contributes to this equation subdominantly in $r$, and so
is not constrained to leading order. (In general, evaluating this
equation to subdominant order in $r$ relates the $n$-dimensional
curvature to the time-evolution of the exponents $\alpha$, $w$ and
$q$.) The leading term vanishes as a consequence of
eq.~\pref{constraint1}, and so does not impose any new conditions.
Neither do the Einstein equations in the $z^a$ directions.

The net summary of the bulk field equations on the parameters $\alpha$,
$w$ and $q$ is therefore given by the two Kasner-like conditions
\pref{constraint1} and \pref{constraint2}. These two conditions
therefore allow a one-parameter family (parameterized, say, by
$q$) of solutions in the vicinity of any given singularity. Notice
that the symmetry of these conditions under $q \to -q$ implies
that for any given asymptotic solution there is a new one which
can be obtained from the first through the weak-to-strong-coupling
replacement $e^\varphi \to e^{-\varphi}$.

Regarding these singularities as brane sources, the one-parameter
set of asymptotic bulk configurations presumably corresponds to a
one-parameter choice which is possible for the couplings of the
brane to bulk fields. For instance, at the lowest-derivative level
considered here this is plausibly related to the choice of dilaton
coupling, such as if the brane action were to take the $D$-brane
form
\be
    S_b = - T \int \exd^n \xi \; \sqrt{-h} \, e^{\lambda \varphi} \,,
\label{dilatoncoupling}
\ee
where $\xi^\mu$ represent coordinates on the brane world-volume,
$T$ is the brane tension, $h_{\mu\nu}$ is the induced metric on
the brane. Here the choice for $\lambda$ (which is a known
function of brane dimension for $D$-branes) plausibly determines
the value of $q$, and so the value of this parameter is not
determined purely from the bulk equations of motion.

We must now go back to ask whether the 
Kasner-like conditions \pref{constraint1} and \pref{constraint2}
are consistent with the requirements $q > -2$ and $q(d-1)-2wn>-2$. 
The first inequality clearly follows from the last of
eqs.~\pref{paramranges}, and so is automatic for the solutions of
interest. By contrast, constraints \pref{constraint1} and
\pref{constraint2} are {\it not} sufficient to ensure that the
second inequality is satisfied, however, as is seen by using
eq.~\pref{constraint1} to rewrite it as $q + 2\,\alpha \ge 0$. This is
clearly not satisfied by the choices $\alpha = 0$, $w = 1/n$ and $q =
-\sqrt{1-1/n}$. Since its violation requires either $\alpha$ or $q$ to
be negative, it necessarily involves either surfaces,
$\Sigma_\alpha$, whose area does not grow with their radius
($\alpha<0$) or the breakdown of the perturbative supergravity
approximation ($q<0$). We exclude such solutions in what follows.

While the requirement $q>-2$ is on solid ground, one might wonder
about the other inequality: $\Xi \equiv q(d-1)-2wn>-2$. In fact, by the 
equations of motion and the assumed asymptotic form for the various fields, the
choice $\Xi < -2$ is not consistent with the requirement that there be a 
nonvanishing flux in the extra dimensions. Similarly, the choice $\Xi = -2$
is also inconsistent if, as before, we require that $\alpha \ge 0$.

\subsection{Asymptotics and Curvature}

We now use the above expressions to evaluate the combination of
bulk fields which appears on the right-hand side of
eq.~\pref{IntegratedCBA}. The surface quantity which appears there
is
\be \label{ICBA-rhs}
    f_\alpha \equiv -\int_{\Sigma_{\alpha}}   d^{d-1} y  \left. \sqrt{\tilde{g}} \,
    {N}_i \left[ W^n \tilde{g}^{ij} \partial_j \left(\,
    \ln W + \frac{2\, \varphi}{D-2} \right) \right]
    \right|_{\Sigma_\alpha}\,,
\ee
and so evaluating this using $N_i \, \exd y^i = -\exd r$ (since
the outward-pointing normal points towards the brane at $r=0$) and
the asymptotic forms given above we find
\be \label{ICBA-rhs-powers}
    \sum_\alpha f_\alpha \sim \sum_\alpha \lim_{r \to 0}
    \left( w_\alpha + \frac{2 \, q_\alpha}{D-2} \right)
    \, c_{\alpha} \, r^{\alpha_\alpha(d-1) + nw_\alpha - 1}
    = \sum_\alpha c_{\alpha} \left( w_\alpha + \frac{2 \, q_\alpha}{D-2} \right)  \,,
\ee
where the last equality uses eq.~\pref{constraint1}. The positive
constants $c_{\alpha}$ are defined by the condition
$\int_{\Sigma_{\alpha}}  d^{d-1} y \sqrt{\tilde{g}} \, W^n \sim
c_{\alpha}\, r^{\alpha_\alpha(d-1) + nw_\alpha }$.

It is the sign (or vanishing) of the sum in
eq.~\pref{ICBA-rhs-powers} which governs the sign (or vanishing)
of the maximally-symmetric $n$-dimensional curvature. Several
points here are noteworthy.
\begin{itemize}
\item $f_\alpha$ always vanishes for any source at which the bulk
equations are nonsingular (or only has a conical singularity),
because $w_\alpha = q_\alpha = 0$ at any such point. Consequently
the maximally-symmetric large $n$ dimensions {\it must} be flat in
the absence of any extra-dimensional brane sources at whose
positions the bulk fields are singular.
\item The $n$-dimensional curvature can vanish even if $f_\alpha
\ne 0$ provided that the sum of the $f_\alpha$'s over all of the
sources vanishes. However such a cancellation requires some of the
$f_\alpha$'s to be negative, and this shows that at there must
exist some sources for which the warping becomes singular
($w_\alpha <0$) or for which the weak-coupling dilaton expansion
fails ($q_\alpha <0$).
\end{itemize}

\section{6D De Sitter solutions}

We next use the above results to construct a new class of
solutions to 6D supergravity which go beyond the known solutions
\cite{GGP,Other6DSolns} by having 4 maximally-symmetric dimensions
which are not flat (see \cite{6DdSnonSUSY} for a recent discussion
of similar solutions in the non-supersymmetric context).

\subsection{Equations of motion}

The action, eq.~\pref{DDAction}, includes as a particular case
that of 6D supergravity coupled to various gauge multiplets
\cite{HiDSugra}, corresponding to the choices $D=6$ and $p_i =
1,2$. In the 6D case $\cA=0$ for ungauged supergravities
\cite{6DSugra}, while $\cA = 2g^2/\kappa^4 \equiv \hat{g}^2/8$ for
chiral gauged supergravity \cite{NS}. For the remainder of this
section we focus on compactifications to 4 dimensions in the
chiral gauged case in the presence of a 2-form flux, $F_{MN}$, for
which $d=2$, $n=4$ and $p=1$.

The equations of motion obtained with these choices are
\ba
    &&\Box \varphi +\frac{\kappa^2}{4} e^{-\varphi} F_{MN}F^{MN}-
    \frac{\kappa^2 \hat{g}^2}{8}e^{\varphi} =0 \\
    &&\nabla_M \( e^{-\varphi} F^{MN} \) =0 \\
    &&R_{MN} + \partial_M \varphi \partial_N \varphi+\kappa^2
    e^{-\varphi}F_{MP}F_{N}^P+\frac{1}{2}(\Box \varphi)g_{MN} =0.
\ea

Following ref.~\cite{GGP} we now make the following \ansatz\ for
the metric
\be
    \exd s^2= \hat{g}_{MN} \, \exd x^M \exd x^N =
    W^2 q_{\mu\nu} \,\exd x^{\mu} \exd x^{\nu}
    +a^2 \exd\theta^2+a^2W^8 \exd\eta^2,
\ee
where the coordinates $(\eta,\theta)$ parameterize the 2 internal
dimensions and $q_{\mu\nu}$ is a maximally-symmetric 4D metric.
(In what follows we take $q_{\mu\nu}$ to be the 4D de Sitter
metric having Hubble constant $H$. The anti-de Sitter case can be
obtained from the final results by taking $H^2 \to - H^2$.) We
assume axial symmetry by requiring $W$, $a$ and $\varphi$ to be
functions only of $\eta$. The gauge potential is taken to have the
monopole form $A = A_{\theta}(\eta) \, \exd\theta$, and so the
only nonzero component of $F$ is $F_{\eta\theta}(\eta)$.

We next write the ordinary differential equations which determine
the unknown functions $\varphi$, $a$ and $W$ and the unknown
constant $H$. To this end, writing the (Maxwell) equation for
$F_{MN}$ as $\partial_M(\sqrt{-g} \, e^{-\varphi} F^{MN}) = 0$
implies $({e^{-\varphi} F_{\eta \theta} }/{a^2})' =0$, where
primes denote differentiation with respect to $\eta$. Integrating
gives
\be
    F_{\eta \theta} = Q \, a^2 e^\varphi,
\ee
where $Q$ is an integration constant, and so in particular
$F_{MN}F^{MN} = 2Q^2 e^{2\varphi}/W^8$.

Using $\hat{\Box}\varphi = \varphi''/(a^2W^8)$ the equation of
motion for the dilaton similarly becomes
\be \label{dilatoneqn}
    \varphi'' + \frac{\kappa^2}{2}Q^2a^2
    e^{\varphi} -\frac{\kappa^2 \hat{g}^2}{8}a^2W^8e^{\varphi}
    =0 .
\ee

Finally, the Einstein equations are obtained using the following
expression for the nonzero components of the Ricci tensor:
\ba
    \hat{R}_{\mu\nu}&=& q_{\mu\nu} \left[ \frac{1}{a^2W^8}
    \Bigl[ WW'' - (W')^2 \Bigr] -  3H^2\,
    \right] \nn\\
    \hat{R}_{\theta\theta}&=&\frac{aa''-(a')^2}{a^2W^8} \\
    \hat{R}_{\eta\eta}&=&\frac{1}{a^2W^2} \Bigl[a W^2 a''
    +4a^2WW'' - W^2 (a')^2
    -8\, aWa' W' -16 \,a^2 (W')^2
    \Bigr] \,. \nn
\ea
Two of the corresponding Einstein equations become
\ba \label{einsteineqn1}
    &&(\mu\nu): \quad
    \frac{W''}{W} - \frac{(W')^2}{W^2}  -  3 \, H^2a^2 W^6
     + \frac12  \, \varphi''
    = 0 \\
    \label{einsteineqn2}
    &&(\theta\theta): \quad
    \frac{a''}{a} - \frac{(a')^2}{a^2} + \kappa^2 Q^2 \, a^2 e^\varphi
    + \frac12 \, \varphi'' = 0
\ea
while use of the $\eta\eta$ component of the Einstein tensor
\be
    \hat{G}_{\eta\eta}= \frac{2}{\, aW^2} \Bigl[ 3 H^2 a^3 W^8 - 2Wa'
    W' -3\, a(W')^2 \Bigr] \,,
\ee
allows the third to be written
\be \label{einsteinconstraint}
    (\eta\eta): \quad 6 \, H^2 a^2 W^6 - \frac{4\,a'
    W'}{aW} - \frac{6(W')^2}{W^2} + \frac12 \, (\varphi')^2 +
    \frac{\kappa^2}{2} Q^2 \, a^2 e^\varphi - \frac{
    \kappa^2 \hat{g}^2}{8} \, a^2 W^8 e^\varphi = 0 \,.
\ee

For numerical purposes we use eqs.~\pref{dilatoneqn},
\pref{einsteineqn1} and \pref{einsteineqn2} to determine
$\varphi''$, $a''$ and $W''$ as a function of $\varphi$, $a$, $W$,
$\varphi'$, $a'$ and $W'$, and by stepping forward in $\eta$
generate a solution as a function of $\eta$. By contrast,
eq.~\pref{einsteinconstraint} must be read as a constraint rather
than an evolution equation because it contains no second
derivatives. The consistency of this constraint with the evolution
equations is guaranteed (as usual) by general covariance and the
Bianchi identities. Evaluating this constraint at the `initial'
point, $\eta = \eta_0$, gives $H$ in terms of the assumed initial
conditions.

\subsection{Solutions}

A general class of solutions to the field equations obtained using
these \ansatze\ is found in ref.~\cite{GGP}, which (using their
conventions for which $\kappa^2 = \frac12$ and $\hat{g}=
4g/\kappa^2 = 8g$) has the form
\bea
    e^\varphi &=& W^{-2} e^{-\lambda_3 \eta} \nonumber\\
    W^4 &=& \left( \frac{Q\lambda_2}{4g\lambda_1} \right)
    \frac{\cosh[ \lambda_1(\eta - \eta_1)]}{\cosh[ \lambda_2
    (\eta - \eta_2)]} \\
    a^{-4} &=& \left( \frac{gQ^3}{\lambda_1^3 \lambda_2}
    \right) e^{-2\lambda_3\eta} \cosh^3[\lambda_1(\eta - \eta_1)]
    \cosh[ \lambda_2(\eta - \eta_2)] \nonumber\\
    \hbox{and} \qquad
    F &=& \left( \frac{Q a^2}{W^2} \right) e^{-\lambda_3
    \eta} \, \exd \eta \wedge \exd \theta \,. \nonumber
\eea
Here $\lambda_i$, $\eta_i$ and $\hat{q}$ are integration
constants, which are subject to the constraint $\lambda_2^2 =
\lambda_1^2 + \lambda_3^2$. For all of these solutions the 4D
metric is flat: $q_{\mu\nu} = \eta_{\mu\nu}$.

These solutions have at most two singularities, and these are
located at $\eta \to \pm \infty$. Locally changing coordinates to
the local proper distance, $\eta \to r_\pm$ with $\exd r_\pm = \mp
a W^4 \, \exd \eta$, brings the singularities at $\eta \to \pm
\infty$ to $r_\pm = 0$, and shows that these solutions have the
asymptotic form described in the previous sections
--- {\it i.e.} eqs.~\pref{dilatonandmetricform} and \pref{Fform}
--- with the powers \cite{GGPplus}
\be
    \alpha_\pm = \frac{\lambda_2 + 3 \lambda_1 \mp 2\lambda_3}{5
    \lambda_2 - \lambda_1 \mp 2 \lambda_3} \,, \quad
    w_\pm = \frac{\lambda_2 - \lambda_1}{5 \lambda_2 - \lambda_1
    \mp 2 \lambda_3}
    \quad \hbox{and} \quad
    q_\pm = -\frac{2( \lambda_2 - \lambda_1 \mp 2 \lambda_3)}{5 \lambda_2 - \lambda_1
    \mp 2 \lambda_3}\,.
\ee
As is easily verified, these satisfy the Kasner-like conditions,
eqs.~\pref{constraint1} and \pref{constraint2}, which for $n=4$
and $d=2$ reduce to $\alpha_\pm + 4 w_\pm = 1$ and $\alpha_\pm^2 + 4 w_\pm^2
+ q_\pm^2 = 1$. As discussed in more detail in
ref.~\cite{GGPplus}, the above expressions imply that the
curvature has a singular limit as $r_\pm \to 0$ unless $\lambda_1
= \lambda_2$ (and so also $\lambda_3 = 0$), in which case these
singularities become conical.

Notice that eq.~\pref{IntegratedCBA} relating brane asymptotics to
the curvature of the 4D space in this case specializes to
\be \label{IntegratedCBA-GGP}
    3 H^2 \int_{-\infty}^\infty \exd\eta \; a^2
    \, W^{6} =
    \left[ \left(\,
    \ln W + \frac{\varphi}{2} \right)'
    \right]_{\eta = -\infty}^{\eta = + \infty} \,.
\ee
Simplifying the right-hand side using the relation $e^\varphi =
W^{-2} e^{-\lambda_3\eta}$ implies $\left( \ln W + \frac12 \varphi
\right)' = - \frac12 \, \lambda_3$, which vanishes only for the
conical-singularity case. However we see that the right-hand side
of eq.~\pref{IntegratedCBA-GGP} nevertheless vanishes once summed
over the two singularities, consistent with the flatness of the 4D
geometries.

\subsection{New Solutions}

We now turn to the construction of more general solutions to the
same field equations, but with the right-hand side of
eq.~\pref{IntegratedCBA-GGP} nonzero and so for which the
maximally-symmetric 4D geometries are not flat. Although we could
do so by directly integrating the field equations as given above,
we instead follow ref.~\cite{GGP} and regard these equations as
coming from the following equivalent Lagrangian
\be
    L=\Bigl[ (\varphi')^2 - 8 (\ln W)' (\ln a)'
    -12 [(\ln W)']^2 \Bigr] N^{-1} - N a^2 e^{\varphi}
    \(\kappa^2 Q^2 -\frac{\kappa^2 \hat{g}^2}{4}W^8+12 H^2 W^6
    e^{-\varphi}\)\,.
\ee
This agrees with the form used in \cite{GGP} when $H = 0$. We
temporarily re-introduce here the `lapse' function,
$g_{\eta\eta}=N^2a^2W^8$, which we may choose coordinates to reset
to unity after it has been varied in the action. Varying with
respect to $N$ gives the constraint equation
(\ref{einsteinconstraint}) where we set $N=1$ after variation.

The equivalent Lagrangian simplifies if we diagonalize the
`kinetic' terms, by defining the new variables $x$, $y$ and $z$
using
\be
    \label{xyz}
    \varphi = \frac{1}{2}(x-y-2z) \,, \quad
    \ln W = \frac{1}{4} (y-x) \quad \hbox{and} \quad
    \ln a = \frac{1}{4}(3x+y+2z) \,.
\ee
In terms of these variables the Lagrangian becomes
\be
    L = (x')^2 - (y')^2 + (z')^2 -
    \kappa^2 Q^2 \, e^{2x} + \frac{\hat{g}^2\kappa^2}{4} \, e^{2y}-12H^2
    \, e^{2y+z}\,.
\ee
We have set $N=1$ but continue to keep in mind its role in
determining the constraint. The `potential' terms simplify further
if we also redefine
\ba
    \label{XYZ}
    X &=& \frac{1}{2} \ln (\kappa^2 Q^2)+x \nn\\
    Y &=& \frac{1}{2} \ln (\hat{g}^2\kappa^2/4)+y \\
    Z &=& \ln (48 |H^2|/\hat{g}^2\kappa^2)+z \nn
\ea
and so
\be
    L = (X')^2 - (Y')^2 + (Z')^2 - e^{2X} + e^{2Y} - \epsilon e^{2Y+Z}
    \,.
\ee where $\epsilon=+1$ for de Sitter and $-1$ for anti-de Sitter
solutions. We now integrate the equations of motion obtained from
this lagrangian to obtain explicit solutions for the
extra-dimensional geometries. Since $X$ has the equation of motion
\be
    X'' + e^{2X} = 0
\ee
it decouples from the other variables. Its equation can be
directly integrated to give
\be
    (X')^2 + e^{2X} = \lambda_1^2,
\ee
and so $e^{-X} = \lambda_1^{-1} \cosh [\lambda_1 (\eta-\eta_1)]$.
The remaining two nontrivial equations of motion become in these
variables
\ba
    \label{Yeom}
    &&Y''+e^{2Y} - \epsilon e^{2Y+Z}=0 \nn\\
    \label{Zeom}
    &&Z'' + \frac{\epsilon}{2} \, e^{2Y+Z} = 0 \,,
\ea
along with the constraint $\lambda_1^2 - (Y')^2 + (Z')^2 -
e^{2Y} + \epsilon e^{2Y+Z}=0$, whose solutions we obtain numerically
below.

In terms of these variables the asymptotic behaviour of the
solutions assumed in previous sections near the singularities is
linear in $\eta$. For example, using eqs.~\pref{xyz} and \pref{XYZ}
to write $X$ in terms of $\varphi$ and $W$, and then using the
asymptotic forms given by eqs.~\pref{dilatonandmetricform} and
\pref{Wform}, we see
\be
    2X = \varphi + 2\ln a + \ln \left(\kappa^2 Q^2 \right)
    \approx     (q_{\pm} + 2\alpha_{\pm} ) \ln r_{\pm}
    \approx  \mp (q_{\pm} + 2\alpha_{\pm} ) \eta,
\ee
where in the last step we have used that $\eta \approx \mp \ln r_{\pm}$ in
the asymptotic region $\eta \rightarrow \pm \infty$. Alternatively,
from the exact solution for $X$ it is clear that
\be
    \lim_{\eta \rightarrow \pm \infty} X \rightarrow
    \mp \lambda_1 \, \eta \,,
\ee
where we take $\lambda_1>0$, corresponding to the condition found
earlier that $(q_{\pm}+2\alpha_{\pm}) \ge 0$. For the other dependent variables
we may similarly write
\ba
    \lim_{\eta \rightarrow \pm \infty} Y &\rightarrow&
    \mp \lambda_2^{\pm} \, \eta, \nn\\
    \lim_{\eta \rightarrow \pm \infty} Z &\rightarrow&
    \mp  \lambda_3^{\pm} \, \eta,
\ea
with independent constants $\lambda_i^\pm$ at $\eta \to \pm
\infty$. By substituting these asymptotic forms into the
differential equations, eqs. \pref{Yeom}, we immediately
obtain the two constraints $\lambda_2^{\pm}>0$ and
$(2\lambda_2^{\pm} + \lambda_3^{\pm})>0$. Note that there is no
restriction on the sign of $\lambda_3^{\pm}$. Finally, the
Kasner-like condition in the asymptotic region also imposes the
following constraint on these constants: $(\lambda_2^{\pm})^2 =
\lambda_1^2 + (\lambda_3^{\pm})^2$.

\begin{figure}
 \begin{center}
 \includegraphics[width=8cm]{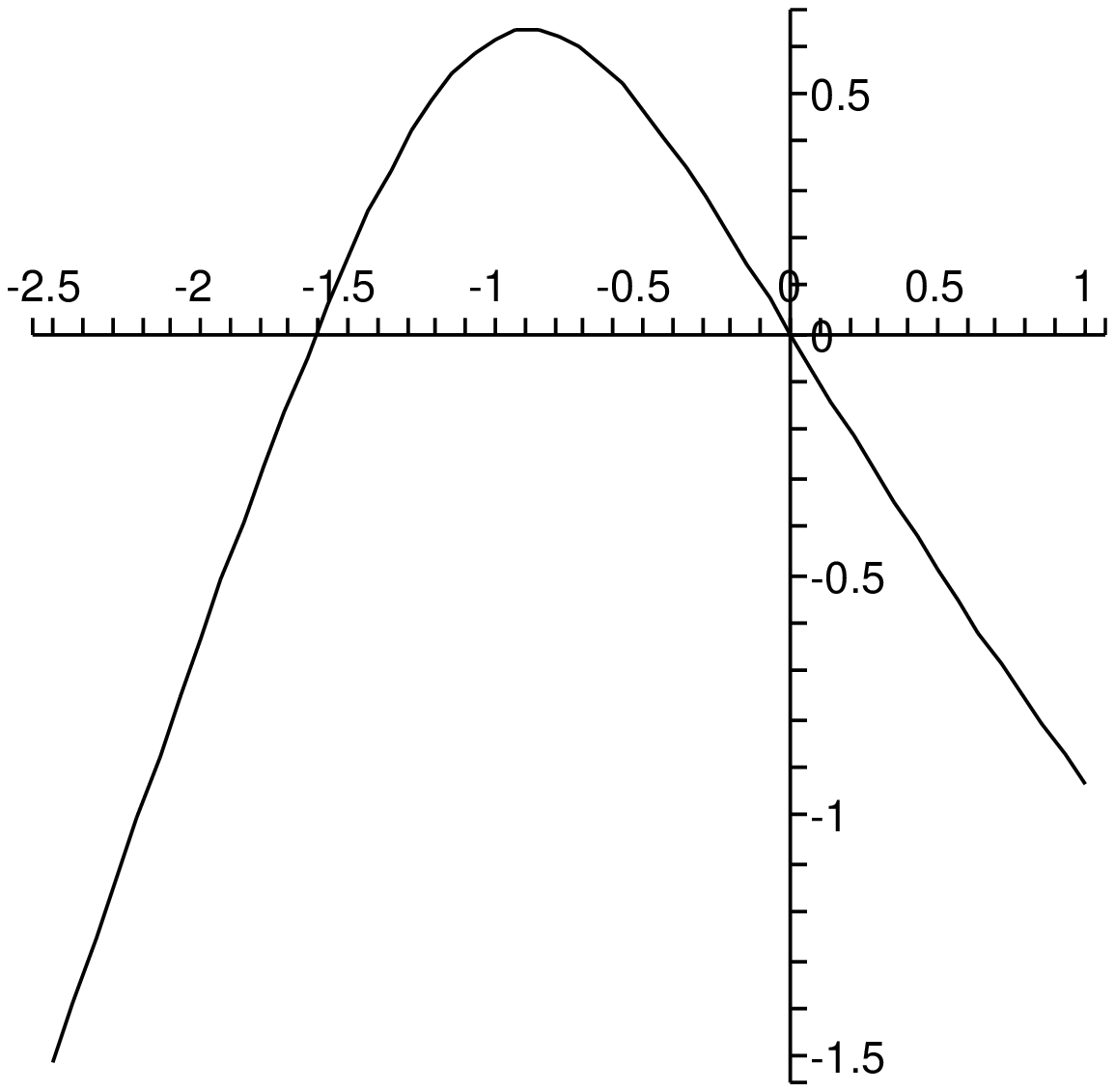}\\
 \caption{Typical behaviour of $Y$ as a function of $\eta$
 for de Sitter solutions ($\epsilon=+1$). The function interpolates
 between two asymptotically linear regimes. The gradient is always
 positive as $\eta \rightarrow -\infty$ and negative as
 $\eta \rightarrow +\infty$.}
 \label{YPlot}
 \includegraphics[width=8cm]{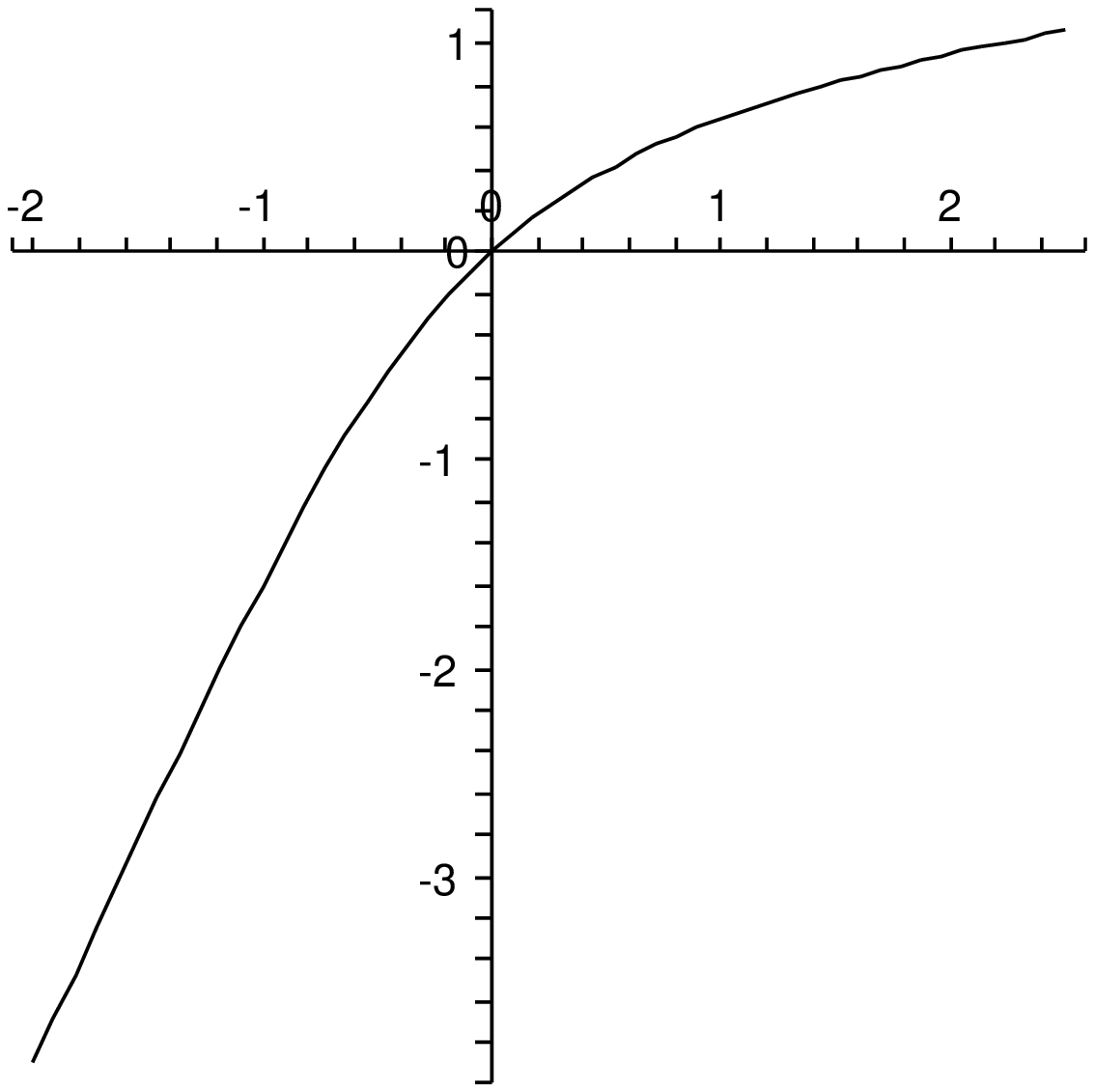}\\
 \caption{Typical behaviour of $Z$ as a function of $\eta$ for
 de Sitter solutions ($\epsilon=+1$). The solutions are asymptotically
 linear with different gradients. For a suitable choice of initial data
 the gradient can change sign as in Fig. \ref{YPlot}.}
 \label{ZPlot}
 \end{center}
 \end{figure}

The solutions of ref.~\cite{GGP} discussed above satisfy these
condition in the special case where $\lambda_3^{\pm}=\pm
|\lambda_3|$, and in this case we know the 4D geometries are flat.
In general, however, both the parameters $\lambda_3^{\pm}$ are not
determined by the one constant $\lambda_3$, and so in the general
case the sum $\sum_\pm f_\pm$ does not vanish, leading ({\it c.f.}
eq.~\pref{IntegratedCBA}) to the conclusion that the corresponding
4D geometries cannot be flat. We have been unable to obtain analytic
solutions to these equations, but there is no obstruction to their
integration. They can be solved numerically leading to numerical
profiles such as those given in figures (\ref{YPlot}) and
(\ref{ZPlot}).

\section{Discussion}

In this paper our focus has been on solutions to the field
equations of the coupled dilaton/$p$-form/Einstein equations of
some $D$-dimensional supergravities, for which $n$ of the
dimensions are maximally symmetric. Such solutions could arise,
for instance, in compactifications from $D$ to $n$ dimensions
within a Kaluza-Klein scenario.

Our main result in this paper is to provide a fairly general
relation between the curvature of the maximally symmetric $n$
dimensions in terms of the (potentially singular) asymptotic
behaviour of the various fields in the vicinity of any brane
sources which may be situated about the internal $d = D-n$
dimensions. This relationship allows an explicit connection to be
made between this curvature and the properties of the branes which
source the geometry. It is only once this connection is made
explicit that it becomes possible to address whether the existence
of flat solutions requires a technically unnatural fine-tuning of
brane properties.

In particular, we use this connection in chiral 6D gauged
supergravity to show the existence of compactifications to 4D de
Sitter and anti-de Sitter geometries having arbitrary curvature.
Since the curvature can be arbitrary it can in particular be
small, implying that these new solutions can be obtained by small
perturbations from the previously-known flat solutions. Since many
of the flat solutions, such as those described by the rugby ball
of ref.~\cite{sled}, are known to have positive tensions it
follows that at least some of these new solutions can also be
sourced by branes having physically reasonable properties.

The existence of such solutions certainly complicates a
self-tuning solution of the cosmological constant problem along
the lines of Ref. \cite{sled} in several ways because it shows
that maximal symmetry in 4 dimensions is insufficient to guarantee
these dimensions must be flat. This means that there are now {\it
two} ways in which perturbations on a brane might destabilize a
flat solution: either by starting a time-dependent runaway or by
generating a maximally-symmetric but curved 4 dimensions. The key
issue which remains is whether the choices of brane properties
which exclude these two options are stable under renormalization.
The results of this paper provide the prerequisite for answering this issue, because they show that the magnitude of the
effective cosmological constant problem is determined by the
asymptotic form of the bulk fields, which are in turn fixed by the
properties of the brane action which we renormalize.

For instance, let us suppose that the branes do not couple to the
dilaton, so that $\lambda=0$ in equation (\ref{dilatoncoupling}),
and suppose that this condition were preserved under quantum
corrections (something which must be explicitly checked). In this
case we anticipate that the boundary condition for the dilaton
near each brane should be $\varphi'=0$, and so $q_{\pm}=0$.
However in such a case the 6D Kasner conditions demand that the
only allowed solutions are those describing conical branes, and
for these our sum-rule implies the only maximally symmetric
solution is Minkowski. Thus if we start out in one of the conical
GGP solutions and perturbation (like renormalization or a phase
transition on one brane) changes the effective brane tension
without growing a dilaton coupling, then the de Sitter and anti-de
Sitter minima discovered here cannot be reached. Furthermore the
system cannot evolve to one of the other non-conical GGP solutions
unless some nontrivial dilaton coupling develops. In such a case
it is likely that the system evolves towards an
as-yet-undiscovered time-dependent runaway solution. If so, the
central question would become how fast the runaway is, and can it
successfully describe the observed Dark Energy? We intend to
return to these questions in a subsequent publication.

\begin{acknowledgments}
We wish to thank Gianmassimo Tasinato for helpful conversations.
This work is supported by a research grant from NSERC (Canada) and
funds from McGill and McMaster Universities and by the Perimeter
Institute. AJT is supported in part by US Department of Energy
Grant DE-FG02-91ER40671.

\end{acknowledgments}

\end{document}